\documentclass[english]{scrartcl}

\usepackage{fullpage}

\usepackage{amstext,amssymb,amsmath,amsthm}
\usepackage{graphicx}
\usepackage{multirow}
\usepackage{url}
\usepackage{hyperref}

\newtheorem{theorem}{Theorem}[section]
\newtheorem{proposition}[theorem]{Proposition}

\title{Approximate inference via variational sampling}

\date{October 7, 2013}

\author{
Alexis Roche\thanks{alexis.roche@gmail.com, alexis.roche@epfl.ch} \\
Siemens Healthcare Sector, Biomedical Imaging Center (CIBM)\\
CH-1015 Lausanne, Switzerland \\
}

\def\x{{\mathbf{x}}}

\def\u{{\mathbf{u}}}

\def\p{{\bar{\mathbf{p}}}}
\def\q{{\bar{\mathbf{q}}}}

\newcommand{\matphi}{\boldsymbol{\Phi}} 

\begin{document}

\maketitle

\begin{abstract}
A new method called ``variational sampling'' is proposed to estimate
integrals under probability distributions that can be evaluated up to
a normalizing constant. The key idea is to fit the target distribution
with an exponential family model by minimizing a strongly consistent
empirical approximation to the Kullback-Leibler divergence computed
using either deterministic or random sampling. It is shown how
variational sampling differs conceptually from both quadrature and
importance sampling and established that, in the case of random
independence sampling, it may have much faster stochastic convergence
than importance sampling under mild conditions. The variational
sampling implementation presented in this paper requires a rough
initial approximation to the target distribution, which may be found,
e.g. using the Laplace method, and is shown to then have the potential
to substantially improve over several existing approximate inference
techniques to estimate moments of order up to two of nearly-Gaussian
distributions, which occur frequently in Bayesian analysis. In
particular, an application of variational sampling to Bayesian
logistic regression in moderate dimension is presented.
\end{abstract}

\section{Introduction}

A central task in probabilistic inference is to evaluate expectations
with respect to the probability distribution of unobserved
variables. Assume that $p:\mathbb{R}^d\to \mathbb{R}_+$ is some
unnormalized target distribution representing, typically, the joint
distribution of some data (omitted from the notation) and some
unobserved $d$-dimensional random vector~$\x$.  Making an inference on
$\x$ often involves computing a vector-valued integral of the form:
\begin{equation}
\label{eq:integral}
I(p) = \int p(\x) \phi(\x) d\x,
\end{equation}
where $\phi:\mathbb{R}^d\to\mathbb{R}^{n}$ is a feature function. For
instance, the posterior mean and variance of $\x$ can be recovered
from (\ref{eq:integral}) if the components of $\phi$ form a basis of
second-order polynomials (including a constant term to capture the
normalizing constant of $p$). A recurring problem, however, is that
(\ref{eq:integral}) may be analytically intractable in practice, so
that one has to resort to approximation methods, which may be roughly
classified as being either sampling-based, or fitting-based, or both.

Sampling-based methods substitute (\ref{eq:integral}) with a finite
sum of the form:
\begin{equation}
\label{eq:integration_rule}
\hat{I}(p) = 
\sum_k w_k \frac{p(\x_k)}{\pi(\x_k)} \phi(\x_k), 
\end{equation}
where $\pi$ is a suitable non-vanishing ``window'' function, while
$\x_k$ and $w_k$ are evaluation points and weights, respectively,
which are determined either deterministically or randomly. In the
deterministic case, (\ref{eq:integration_rule}) is called a quadrature
rule. Classical Gaussian quadrature is exact, by construction, if the
integrand $(p/\pi) \phi$ is polynomial in each component. Other rules
known as Bayes-Hermite quadrature model the integrand as a spline or,
equivalently, as the outcome of a Gaussian process \cite{OHagan-91}.
For computational efficiency in multiple dimensions, quadrature rules
should be implemented using sparse grids \cite{Gerstner-98}, a basic
example of which is used, e.g., in the unscented Kalman filter
\cite{Julier-04}.

An alternative to quadrature is to sample the evaluation points
randomly and independently from the window function $\pi$ considered
as a probability density, leading to the popular importance sampling
(IS) method \cite{Metropolis-87}. Setting uniform weights $w_k=1/N$,
where $N$ is the sample size, then guarantees that
(\ref{eq:integration_rule}) yields a statistically unbiased estimate
of $I(p)$. This procedure, however, may have slow stochastic
convergence, or may not converge at all, if the integrand $(p/\pi)
\phi$ has large or infinite variance under $\pi$. The most common
approach to mitigate this problem is to constrain sampling so as to
reduce fluctuations of the integrand across sampled points. This has
motivated an impressive number of IS enhancements proposed in the
literature, including defensive and multiple IS \cite{Owen-00},
adaptive IS \cite{Bucher-88,Cappe-04}, path sampling \cite{Gelman-98},
annealed IS \cite{Neal-01}, sequential Monte Carlo methods
\cite{DelMoral-06}, Markov chain Monte Carlo (MCMC) methods
\cite{Andrieu-03,Andrieu-08,Girolami-11}, and nested sampling
\cite{Skilling-06}.

In contrast with sampling-based methods, fitting-based methods
evaluate (\ref{eq:integral}) by replacing the target distribution~$p$
with an approximation for which the integral is tractable, thereby
usually avoiding sampling. For instance, the well-known Laplace method
\cite{Tierney-86} performs a Gaussian approximation using the
second-order Taylor expansion of $\log p$ at the mode. In some
problems, however, more accurate Gaussian approximations may be
obtained by iterative analytical calculation using variational methods
such as lower bound maximization
\cite{Opper-09,Seeger-10,Hall-11,Ormerod-12} or expectation
propagation (EP) \cite{Minka-01,Minka-05}. Other types of explicit
approximations (not necessarily Gaussian) may be computed using a
latent variable model and applying the variational Bayes algorithm
\cite{Beal-03} or using other structured variational inference methods
\cite{Jordan-99,Minka-05}.

Fitting-based methods, however, are not always applicable as they
typically rely on parametric or structural assumptions regarding the
target distribution. When applicable, they may be computationally
efficient but may lack accuracy since they intrinsically rely on
approximations.  On the other hand, sampling-based methods are widely
applicable and virtually not limited in accuracy but tend to be
time-consuming in practice in that they may require many function
evaluations to converge.

This paper explores a third group of methods that combine sampling and
fitting ideas. As in variational methods, the evaluation of
(\ref{eq:integral}) is recast into an optimization problem. However,
rather than choosing the objective function for analytical
tractability, the idea is to approximate, by sampling, an intractable
objective whose optimization is known to yield an {\em exact}
result. The rational behind this {\em variational sampling} idea is
that a well-chosen variational formulation may be statistically more
efficient than a direct integral evaluation as in
(\ref{eq:integration_rule}). While previous methods in this vein used
weighted log-likelihood as a sampling-based approximation to the
Kullback-Leibler (KL) divergence
\cite{Wei-90,DeBoer-05,Carbonetto-09}, this paper advocates another
very natural way to approximate the KL~divergence that yields more
accurate integral estimates under mild conditions and, perhaps
surprisingly, does not seem to have been proposed before.

\section{Variational sampling}
\label{sec:variational_sampling}

Our starting point is a well-known variational argument: computing the
integral (\ref{eq:integral}) is equivalent, under some existence
conditions, to substituting $p$ with the distribution $q_\star$ that
minimizes the inclusive KL divergence:
\begin{equation}
\label{eq:kl_div}
L_\star(q) = D(p\|q) 
= \int p(\x) \log \frac{p(\x)}{q(\x)}d\x 
- \int p(\x)d\x
+ \int q(\x)d\x,
\end{equation} 
over the exponential family:
\begin{equation}
\label{eq:exp_family}
q_\theta (\x) = e^{\theta^\top \phi(\x)},
\end{equation}
which is induced by the feature function~$\phi$. Here, we consider the
generalized KL~divergence as, e.g. in \cite{Minka-05}, which is a
positive-valued convex function of $(p,q)$ for any pair of {\em
  unnormalized} distributions, and vanishes iff $p=q$. Remarkably,
since $I(q_\star)=I(p)$, the variational approximation $q_\star$
yields an error-free integral for this particular choice of divergence
and approximating family.  We note that $I(q_\star)$ has a closed-form
expression if the exponential family consists of Gaussian or finite
distributions, or products of convenient univariate distributions. The
experiments reported in Section~\ref{sec:experiments} focus on the
Gaussian family, where $\phi$ spans the space of second-order
polynomials in $\x$, in which case $q_\star$ is the Gaussian
distribution with same normalizing constant, mean and variance matrix
as $p$.

Note that, since we have the decomposition
$D(p\|q_\theta)=D(p\|q_\star)+D(q_\star\|q_\theta)$ for any $q_\theta$
in the exponential family, minimizing (\ref{eq:kl_div}) boils down to
minimizing the {\em excess} KL divergence $D(q_\star\|q_\theta)$,
which is clearly a positive quantity that vanishes iff $q_\theta=
q_\star$. Provided that $\theta$ defines a unique parametrization, the
excess KL divergence is nonzero whenever $I(q_\theta)$ differs from
$I(q_\star)=I(p)$ and therefore it defines a natural error measure on
the integral (\ref{eq:integral}), which we will use in
Section~\ref{sec:experiments} to assess estimation accuracy.

\subsection{Surrogate KL divergence}

\begin{figure}[!ht]
  \begin{center}
    \includegraphics[width=0.4\textwidth]{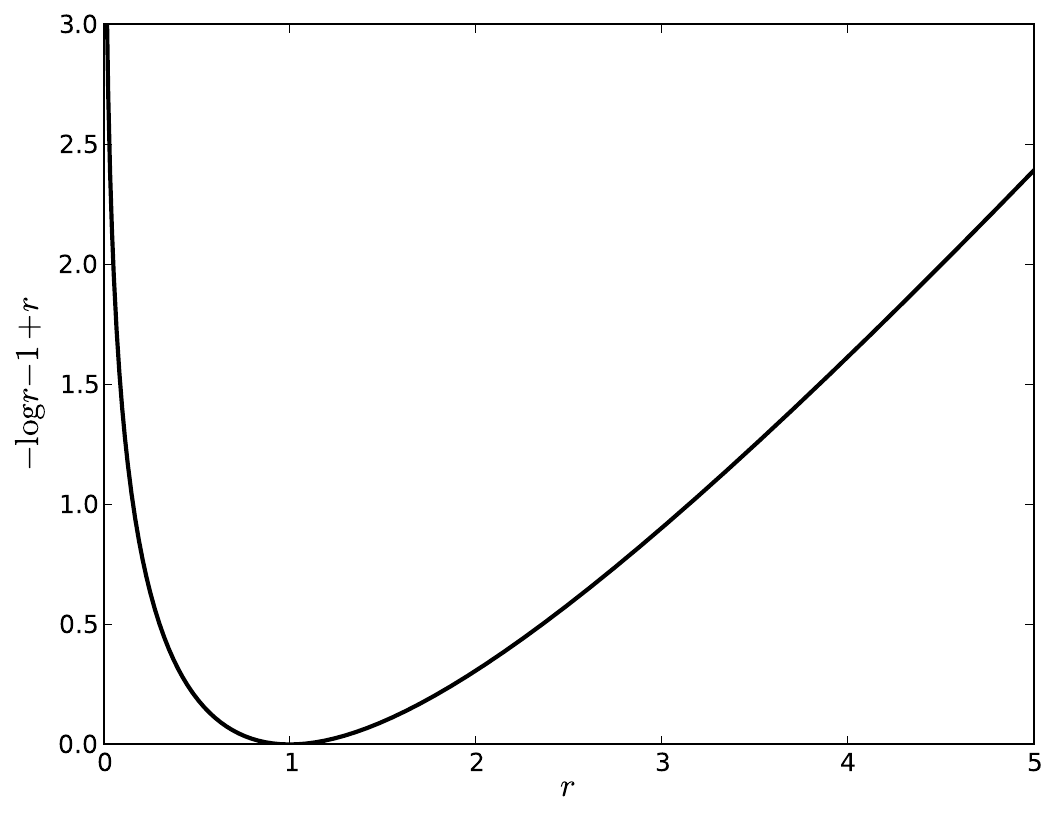}
  \end{center}
\caption{Loss function associated with the generalized KL divergence.}
\label{fig:kl_loss}
\end{figure}

Unfortunately, evaluating (\ref{eq:kl_div}) for any~$q$ and thus
finding $q_\star$ is just as difficult as evaluating
(\ref{eq:integral}). However, it seems natural to use a sampling
approximation of $L_\star(q)$ as a surrogate cost function to be
minimized. To this end, we remark that $L_\star(q)$ is the expected
value under $p$ of the loss function $\ell(\x, q) = - \log(r(\x)) - 1
+ r(\x)$ with $r(\x) \equiv {q(\x)}/{p(\x)}$. Applying an integration
rule such as~(\ref{eq:integration_rule}) to $\ell(\x, q)$, we may
propose the following KL approximation:
\begin{equation}
\label{eq:vs_cost}
L(q) = 
\sum_k 
w_k
\Big[
\frac{p(\x_k)}{\pi(\x_k)}
\log \frac{p(\x_k)}{q(\x_k)}
- \frac{p(\x_k)}{\pi(\x_k)}
+ \frac{q(\x_k)}{\pi(\x_k)}
\Big].
\end{equation}
As depicted in Figure~\ref{fig:kl_loss}, the loss $\ell(\x,q)$ at any
point $\x$ is minimized and vanishes iff $r(\x)=1$, meaning
$q(\x)=p(\x)$. This implies the {\em strong consistency} property that
$L(p)\leq L(q)$ for any unnormalized distribution $q$ possibly outside
the approximating exponential family, with equality iff $q$ coincides
with $p$ at the sampled points. In other words, $L$ defines a valid
cost function not just in an asymptotic sense, but for any finite
sample. Another way to see this is to remark that (\ref{eq:vs_cost})
may be interpreted as a {\em finite} KL~divergence $D(\p\|\q)$, where
the weighted sample vectors:
\begin{equation}
\label{eq:discrete_kl}
\p = \Big[w_1\frac{p(\x_1)}{\pi(\x_1)}, 
w_2\frac{p(\x_2)}{\pi(\x_2)}, 
\ldots
\Big]^\top,
\quad
\q = 
\Big[w_1\frac{q_\theta(\x_1)}{\pi(\x_1)}, 
w_2\frac{q_\theta(\x_2)}{\pi(\x_2)}, 
\ldots
\Big]^\top
,
\end{equation}
are considered as unnormalized finite distributions. This remark
suggests that our variational sampling strategy essentially recasts a
continuous distribution fitting problem into a discrete one.

We note that another possible sampling-based approximation of
(\ref{eq:kl_div}) is:
\begin{equation}
\label{eq:is_cost}
L_0(q) = 
\sum_k 
w_k
\frac{p(\x_k)}{\pi(\x_k)}
\Big[
\log \frac{p(\x_k)}{q(\x_k)}
- 1 
\Big]
+
\int q(\x) d\x
,
\end{equation}
which slightly differs from (\ref{eq:vs_cost}) in that the sampling
approximation to the integral of $q$ is replaced by its exact
expression. Minimizing (\ref{eq:is_cost}) over the exponential family
is akin to a familiar maximum likelihood estimation problem, yielding
the moment-matching condition:
$$
\int q_\theta(\x) \phi(\x) d\x
=
\sum_k w_k \frac{p(\x_k)}{\pi(\x_k)} \phi(\x_k).
$$ In other words, minimizing $L_0$ provides the same integral
approximation as a direct application of the integration rule to $p$,
and has therefore no practical value in the context of estimating
integrals. $L_0$ has been used previously by analogy with maximum
likelihood estimation for various distribution approximation tasks,
e.g. in the Monte Carlo expectation-maximization algorithm
\cite{Wei-90} or in the cross-entropy method
\cite{DeBoer-05}. 

\begin{figure}[!ht]
  \begin{center}
    \includegraphics[width=.49\textwidth]{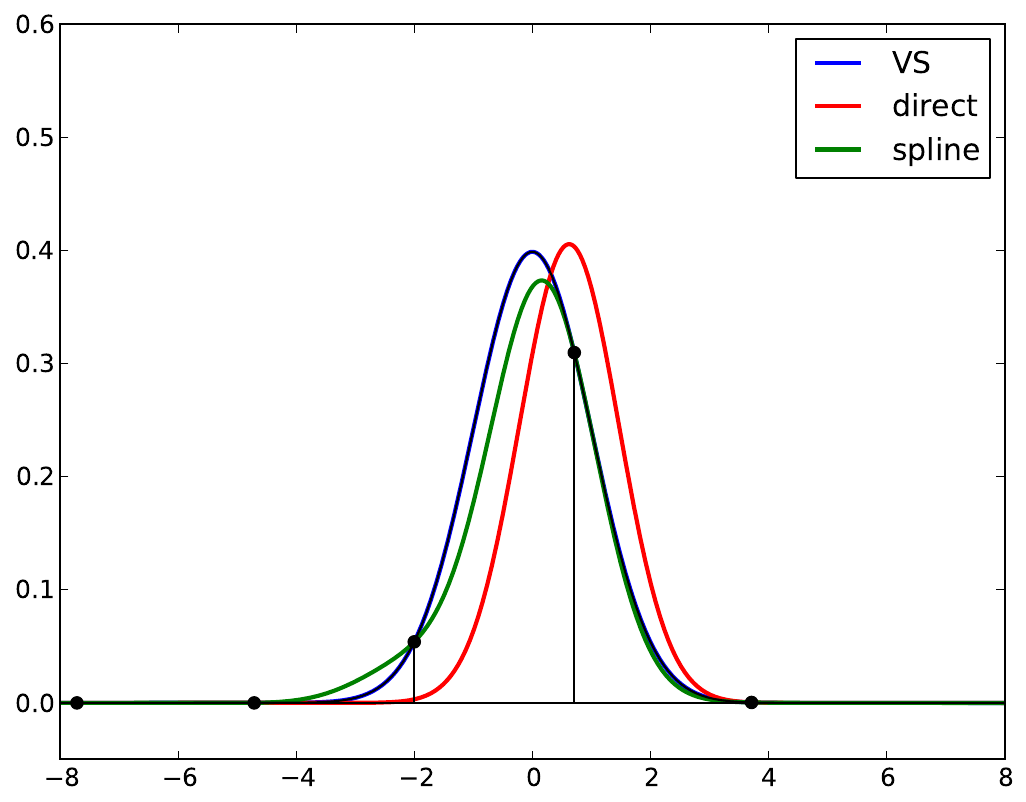}
    \includegraphics[width=.49\textwidth]{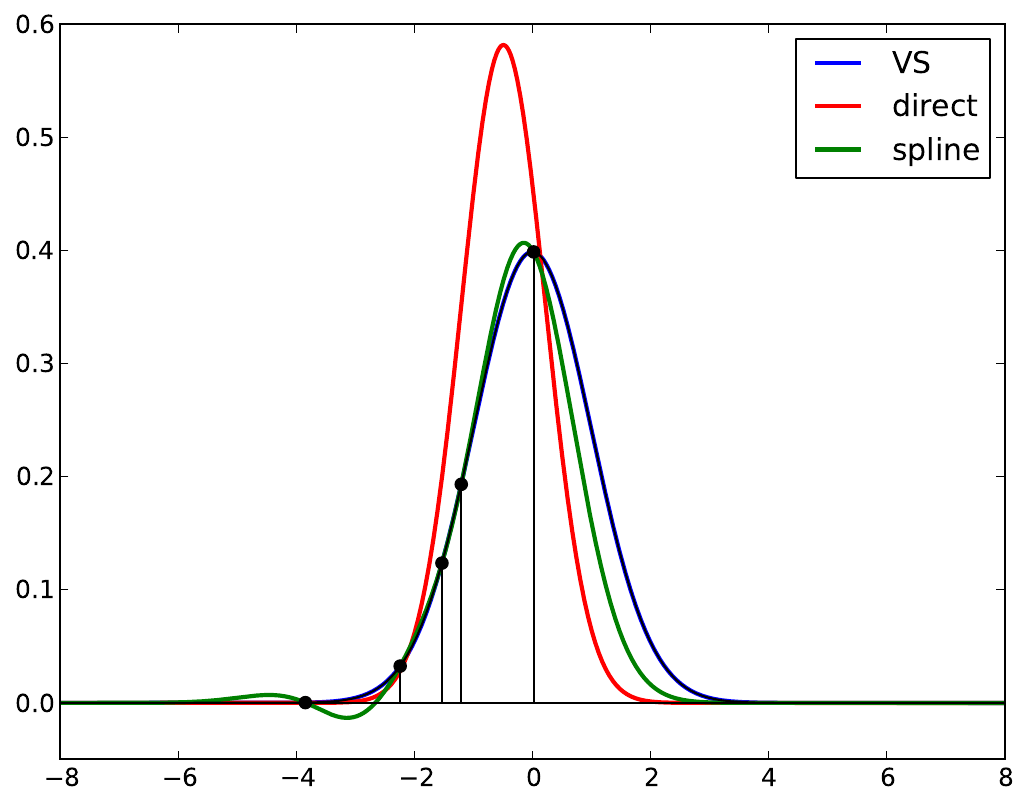}
  \end{center}
  \caption{Exactness of variational sampling for fitting a normal
    distribution $N(0,1)$ using a Gauss-Hermite rule (left) and random
    independent sampling (right) both with window function
    $\pi=N(-2,4)$. Blue lines: target distribution and VS Gaussian
    fits. Red lines: Gaussian fits obtained by direct integral
    approximations (equivalent to IS in the random sampling
    case). Green lines: interpolating Gaussian splines.}
  \label{fig:exactness}
\end{figure}

However, $L_0$ cannot be interpreted as a discrete divergence,
contrary to $L$, and turns out not to satisfy the above-mentioned
strong consistency property of $L$ to be globally minimized by~$p$,
meaning that it is generally possible to find a distribution~$q$, even
in a restricted approximation space, such that $L_0(q)<L(p)$, as
illustrated in Figure~\ref{fig:exactness}. In the sequel, we refer to
variational sampling (VS) as the minimization of $L$ as defined in
(\ref{eq:vs_cost}). Owing to the strong consistency of $L$ as a KL
divergence approximation, VS may produce more accurate integral
estimates than a direct application of the integration rule if the
target distribution is ``close enough'' to the fitting family.

\subsection{Minimization}

When choosing the approximation space as the exponential family
(\ref{eq:exp_family}), $L(q_\theta)$ is seen to be a smoothly convex
function of $\theta$, abbreviated $L(\theta)$ in the following, with
gradient and Hessian matrix:
$$
\nabla_\theta L
=
\matphi^\top(\q_\theta-\p),
\qquad 
\nabla_\theta \nabla_\theta^\top L
= \matphi^\top {\rm diag}(\q_\theta) \matphi,
$$ where $\p$ and $\q_\theta$ are the weighted samples of $p$ and $q$
as defined in (\ref{eq:discrete_kl}), and $\matphi$ is the $N\times n$
matrix with general element $\matphi_{kj}=\phi_j(\x_k)$. The Hessian
may be interpreted as a Gram matrix and is thus positive definite
everywhere provided that $\matphi$ has rank $n$. While there is
generally no closed-form solution to the minimization
of~(\ref{eq:vs_cost}), the following result provides a simple
sufficient condition for the existence and uniqueness of a minimum.

\begin{proposition}
\label{prop:uniqueness}
If the matrix $\matphi$ has rank $n$, then $L(\theta)$ admits a unique
minimizer in $\mathbb{R}^n$.
\end{proposition}

\begin{proof}[Proof]
If $\matphi$ has rank $n$, then $\matphi^\top{\rm
  diag}(\q_\theta)\matphi$ is positive definite for any~$\theta$. This
implies that $L(\theta)$ is strictly convex and guarantees
uniqueness. Existence then follows from the fact that $L(\theta)$ is
coercive in the sense that $\lim_{\|\theta\|\to\infty}
L(\theta)=+\infty$. To prove that, we decompose $\theta$ via $\theta =
t \u$ where $t$ is a scalar and $\u$ a fixed unit-norm
vector. $\matphi$ having rank~$n$ means that $\u^\top \matphi \not=0$,
hence there exists at least one~$k$ for which $\sum_i u_i
\phi_i(\x_k)\not=0$, implying that $\lim_{t\to+\infty}
q_{t\u}(\x_k)/p(\x_k) \in \{0,+\infty\}$ therefore $\lim_{t\to+\infty}
\ell(\x_k, q_{t\u})=+\infty$, which completes the proof.
\end{proof}

For many choices of $\phi$ functions, such as linearly independent
polynomials, this condition is verified as soon as $N\geq n$ distinct
points are sampled. For instance, the minimum sample size to guarantee
uniqueness when fitting a full Gaussian model is $n=(d+2)(d+1)/2$.

In small dimension, say $d<10$, we perform numerical optimization
using the Newton method with adaptive step size, which involves
inverting the Hessian matrix at each iteration using a Cholesky
decomposition. This step is time consuming in larger dimension, hence
we resort to a quasi-Newton approach where the Hessian is approximated
by the fixed matrix $\matphi^\top {\rm diag}(\p) \matphi$, which gives
a rough estimate of the Hessian at the minimum. Using this trick, the
optimizer typically requires more iterations to converge but runs
faster overall because iterations are much cheaper.

\section{Monte Carlo implementation}
\label{sec:monte_carlo}

So far, we have considered a general definition of variational
sampling that relies on any deterministic or stochastic integration
rule of type~(\ref{eq:integration_rule}). In practice, it is simple to
implement the IS rule, where points are sampled independently from a
probability density $\pi$ and weights are uniform, $w_k=1/N$. In this
context, VS is a Monte Carlo integration method comparable to IS, and
enjoys an analogous central limit theorem that states, in short, that
the VS integral estimator is asymptotically unbiased with variance
inversely proportional to $N$.

\begin{proposition}
\label{prop:central_limit}
Under the conditions of Proposition~\ref{prop:uniqueness}, let
$\hat{\theta}=\arg\min_{\theta\in\mathbb{R}^n} L(\theta)$ and let the
corresponding integral estimator $\hat{I}(p)=I(q_{\hat{\theta}})$. If
$D_\star(p\|q)$ admits a unique minimizer $q_\star$ in the exponential
family, then $\hat{I}(p)$ converges in distribution to $I(p)$:
\begin{equation}
\label{eq:asymptotic_variance}
\sqrt{N}[\hat{I}_p(\phi)-I_p(\phi)] \stackrel{d}{\to} {\cal N}(0, \Sigma), 
\qquad {\rm with} \quad
\Sigma = \int
\frac{[p(\x)-q_\star(\x)]^2}{\pi(\x)}\phi(\x)\phi(\x)^\top
d\x
,
\end{equation}
provided that $\Sigma$ exists and is finite.
\end{proposition}

\begin{proof}[Sketch of proof]
The proof is based on standard Taylor expansion-based arguments from
asymptotic theory and is straightforward using a general convergence
result on Monte Carlo approximated expected loss minimization given by
Shao \cite{Shao-89}, theorem~3. This theorem implies under the stated
conditions that $\hat{\theta}$ converges in distribution
to~$\theta_\star$. The result then easily follows from a Taylor
expansion of $I(q_{\theta})$ around $\theta_\star$.
\end{proof}

As corollaries of Proposition~\ref{prop:central_limit}, we also have
that $\sqrt{N}(\hat{\theta}-\theta_\star)$ converges in distribution
to $N(0,H^{-1}\Sigma H^{-1})$, where $H$ is the Hessian of the
continuous KL~divergence at the minimum: $H = \int
q_\star(\x)\phi(\x)\phi(\x)^\top d\x$, and that the minimum
KL~divergence is overestimated by a vanishing portion:
$$
L_\star(\hat{\theta}) - L_\star(\theta_\star)
= D(q_\star\|q_{\hat{\theta}}) = 
\frac{1}{2N} {\rm
  trace}(\Sigma H^{-1}) + o(\frac{1}{N}),
$$ meaning that the integral estimation error, as measured by the
excess KL~divergence, also decreases in $1/N$.

The VS asymptotic variance $\Sigma/N$ is to be compared with that of
the IS estimator, which is given by $\Sigma_0/N$ with:
\begin{equation}
\label{eq:is_variance}
\Sigma_0
= \int \frac{p(\x)^2}{\pi(\x)} \phi(\x)\phi(\x)^\top d\x
- I(p)I(p)^\top.
\end{equation}

Both variances decrease in $O(1/N)$ yet with factors that might differ
by orders of magnitude if the function $p-q_\star$ takes on small
values. Substituting $p$ with $p-q_\star$ in (\ref{eq:is_variance}),
we get the fundamental insight that VS is asymptotically equivalent to
IS applied to $p-q_\star$ {\em even though $q_\star$ is unknown}. An
interpretation of this result is that VS tries to compensate for
fluctuations of the importance weighting function, $p/\pi$ by
subtracting it a best fit (in the KL~divergence sense), $q_\star/\pi$.
We retrieve the fact that VS is exact in the limiting case
$p=q_\star$.  In this light, VS appears complementary to IS extensions
that reduce variance by constraining the sampling mechanism
\cite{Bucher-88,Gelman-98,Owen-00,Cappe-04,Neal-01,DelMoral-06}.

From Proposition~\ref{prop:central_limit}, a sufficient condition for
stochastic convergence is that $\Sigma$ be finite. Loosely speaking,
this means that $\pi$ should be ``wide enough'' to sample regions
where $p$ and $q_\star$ differ significantly. In practice, choosing
$\pi$ as an initial guess of $q_\star$ found, e.g. using the Laplace
method, is often good enough to get fast convergence. This strategy,
however, does not offer a strict warranty. In cases where $p$ or
$q_\star$ have significantly stronger tails than the chosen $\pi$,
convergence might not occur or be too slow. While such cases may be
diagnosed using an empirical estimate of $\Sigma$, they will
necessitate to restart the method, e.g. by rescaling the variance of
$\pi$. A strategy that alleviates the need for such a calibration
step, is to pre-multiply $p$ by a fast decreasing ``context'' function
$c$ and search for a fit of the form $c(\x)q_\theta(\x)$. In this
straightforward extension, VS minimizes a sampling approximation to
the localized KL~divergence $D_c(p\|q_\theta)=D(cp\|cq_\theta)$ as
opposed to the global divergence $D(p\|q_\theta)$, and faster
convergence then comes at the price of a biased integral
estimate. This approach may be viewed as a tradeoff between the
Laplace method and global KL~divergence minimization.

\section{Experiments}
\label{sec:experiments}

We implemented the Monte Carlo-based VS method described in
Sections~\ref{sec:variational_sampling} and \ref{sec:monte_carlo} in
Scientific Python (\url{www.scipy.org}). The code is in open-source
access at \url{https://github.com/alexis-roche/variational_sampler}.
The following describes some experiments using VS to illustrate its
practical value compared to other methods.

\subsection{Simulations}
\label{sec:simulations}

We simulated non-Gaussian target distributions as mixtures of 100
normal kernels with random centers and unit variance:
\begin{equation}
\label{eq:mixture}
p_{\delta}(\x) = \frac{1}{100} \sum_{i=1}^{100} N(\x; \mu_i, \mathbf{I}_d),
\qquad
\mu_i \stackrel{i.i.d.}{\sim} N(0, \frac{\delta^2}{d}\mathbf{I}_d).
\end{equation}
The parameter $\delta^2$ represents the average squared Euclidean
distance between centers, and thus controls deviation of $p_\delta$
from normality.

\begin{figure}[!ht]
  \begin{center}
    \includegraphics[width=.49\textwidth]{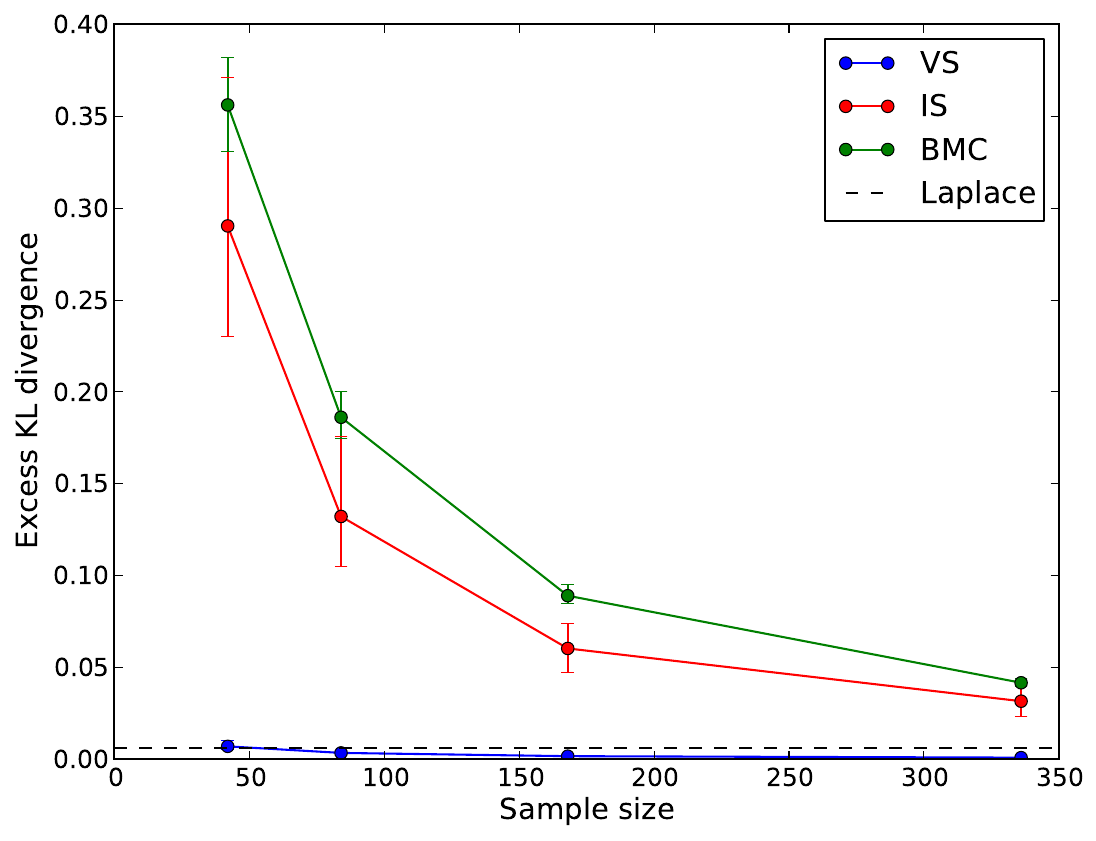}
    \includegraphics[width=.49\textwidth]{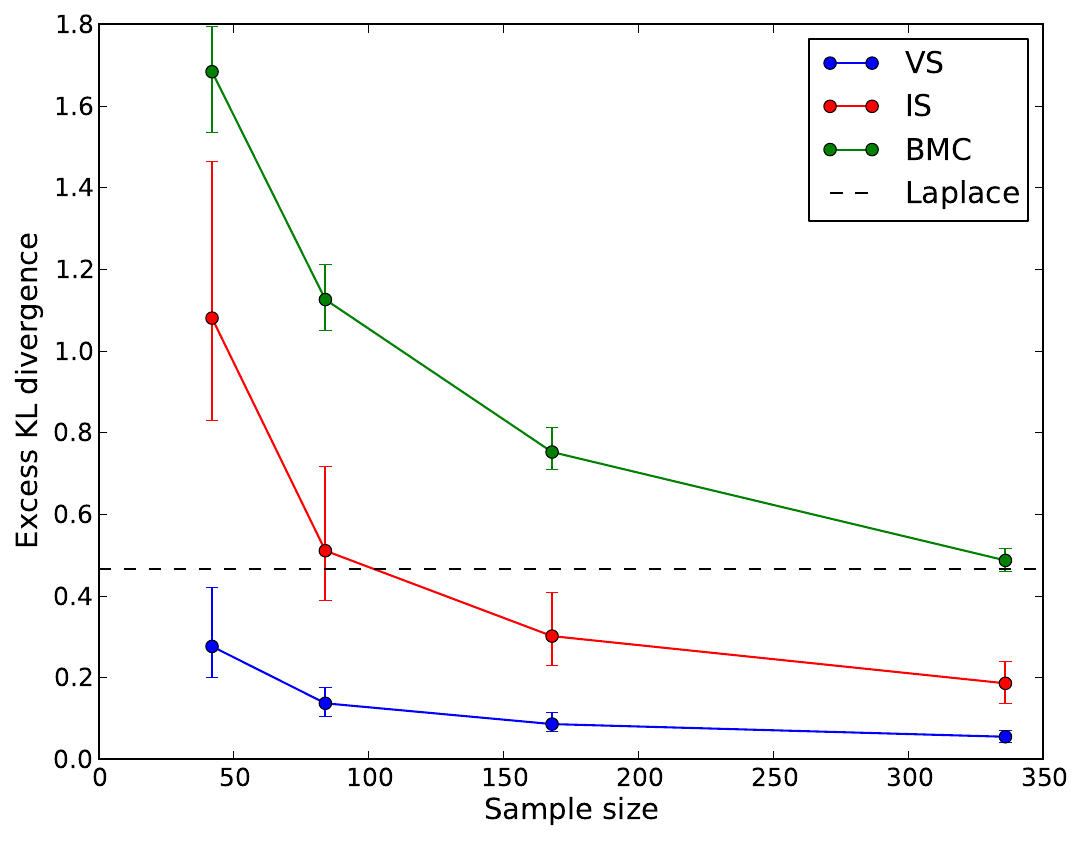}
    \includegraphics[width=.49\textwidth]{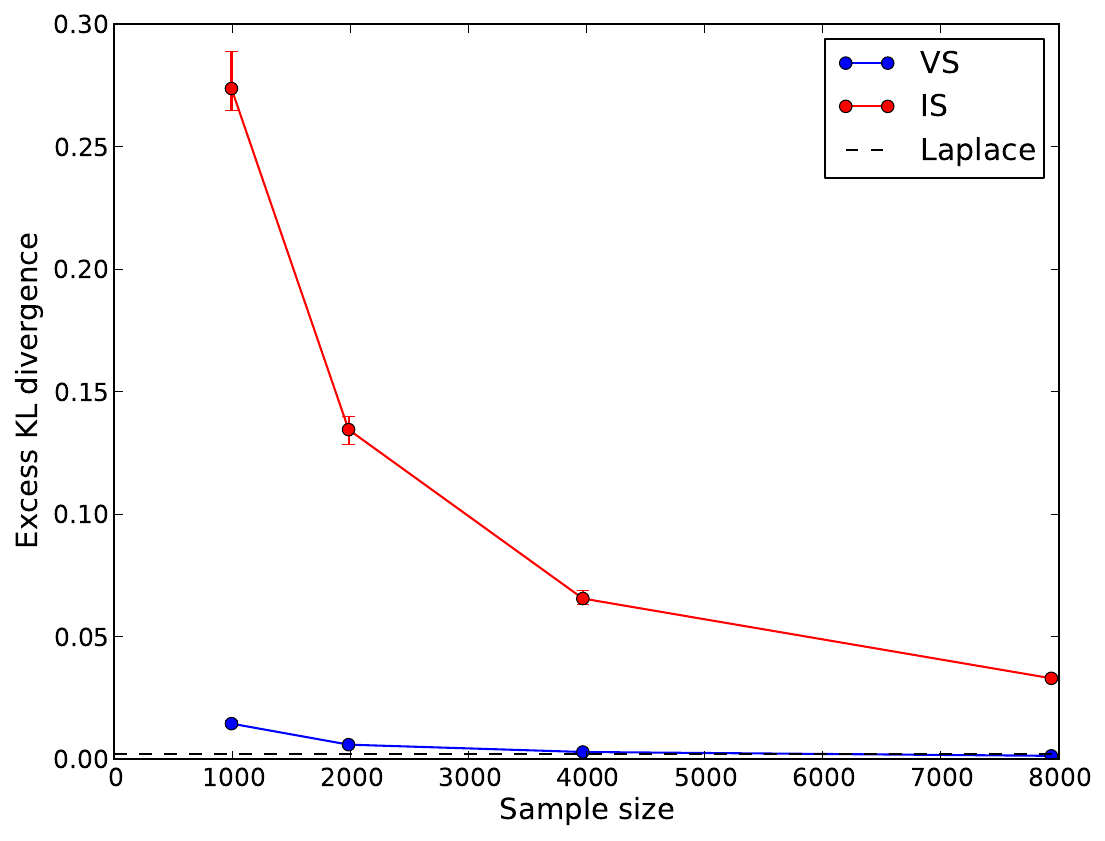}
    \includegraphics[width=.49\textwidth]{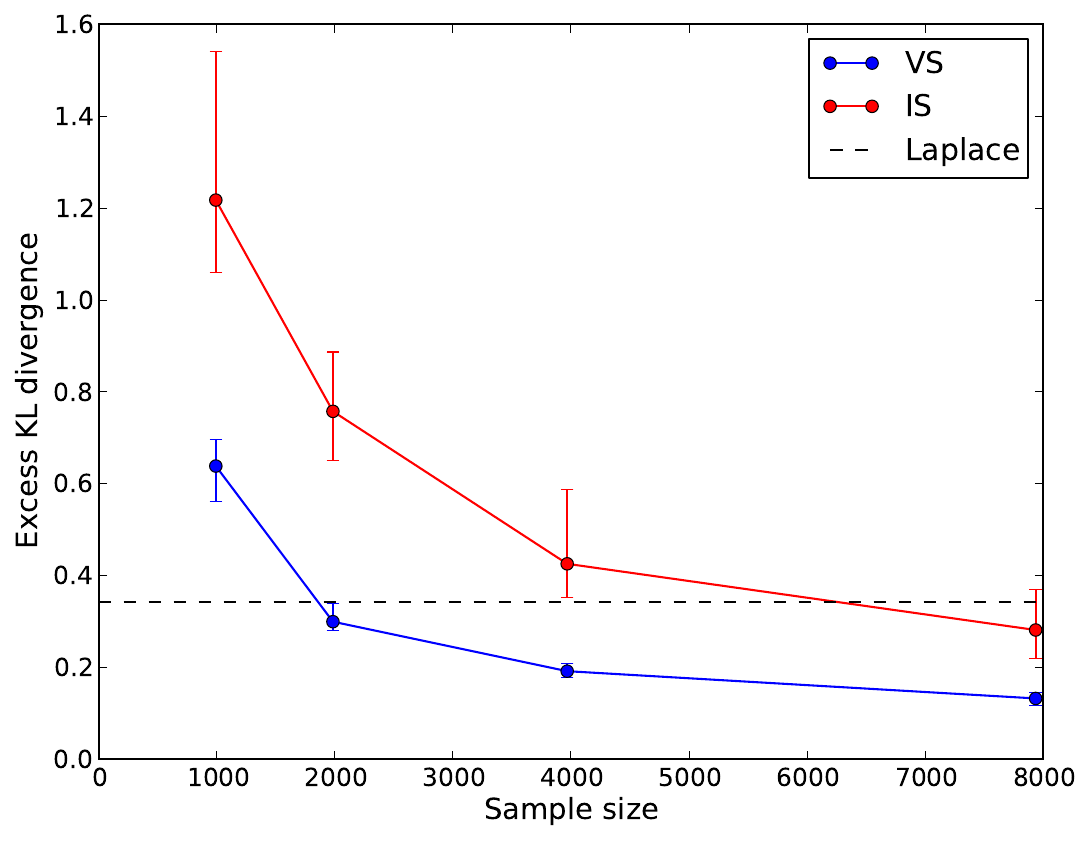}
  \end{center}
  \caption{Excess KL~divergences of VS, IS, BMC and the Laplace
    methods for simulated Gaussian mixture target distributions as in
    (\ref{eq:mixture}). Left, $\delta=1.5$. Right, $\delta=3$. Top, in
    dimension~5. Bottom, in dimension~30. BMC errors in dimension~30
    are out of the represented range. Dots and vertical segments
    respectively represent median errors and inter-quartile ranges
    across 100 sampling trials.}
  \label{fig:simulations}
\end{figure}

Figure~\ref{fig:simulations} compares performances of VS, IS, Bayesian
Monte Carlo (BMC) \cite{Rasmussen-03} and the Laplace method in
estimating the $n=(d+2)(d+1)/2$ moments of order 0, 1 and 2 of
$p_\delta$, in different dimensions, for different values of $\delta$
and for different sample sizes. VS, IS and BMC were run on the same
random samples drawn from the Laplace approximation to $p_\delta$. VS
was implemented to fit a full Gaussian model to $p_\delta$, as
described in Section~\ref{sec:variational_sampling}. The BMC method
works by fitting an interpolating spline and was implemented here
using an isotropic Gaussian correlation function with fixed scale~1
that matches the Gaussian kernels in $p_\delta$.

Errors are measured by the excess KL~divergence $D(q_\star \|
\hat{q})$ (see Section~\ref{sec:variational_sampling}), where
$\hat{q}$ is the Gaussian fit output by a method and $q_\star$ is the
known KL~optimal fit analytically derived from $p_\delta$. Note that
$D(q_\star \| \hat{q})$ can be calculated analytically as it is the
KL~divergence between two unnormalized Gaussian distributions. It is a
global error measure that combines errors on moments of order 0, 1 and
2, hence it departs from zero whenever either the normalizing
constant, or the posterior mean, or the posterior variance matrix is
off.
 
These simulations illustrate the accelerated stochastic convergence of
VS over IS for target distributions that are close to Gaussian. For
almost Gaussian distributions, VS is merely equivalent to the Laplace
method. As the target departs from normality, VS can significantly
improve over the Laplace method while still requiring much smaller
samples than IS for comparable accuracy. The BMC method was here
competitive with IS in dimension~5, but broke down in higher
dimension.

\begin{figure}[!ht]
  \begin{center}
    \includegraphics[width=.5\textwidth]{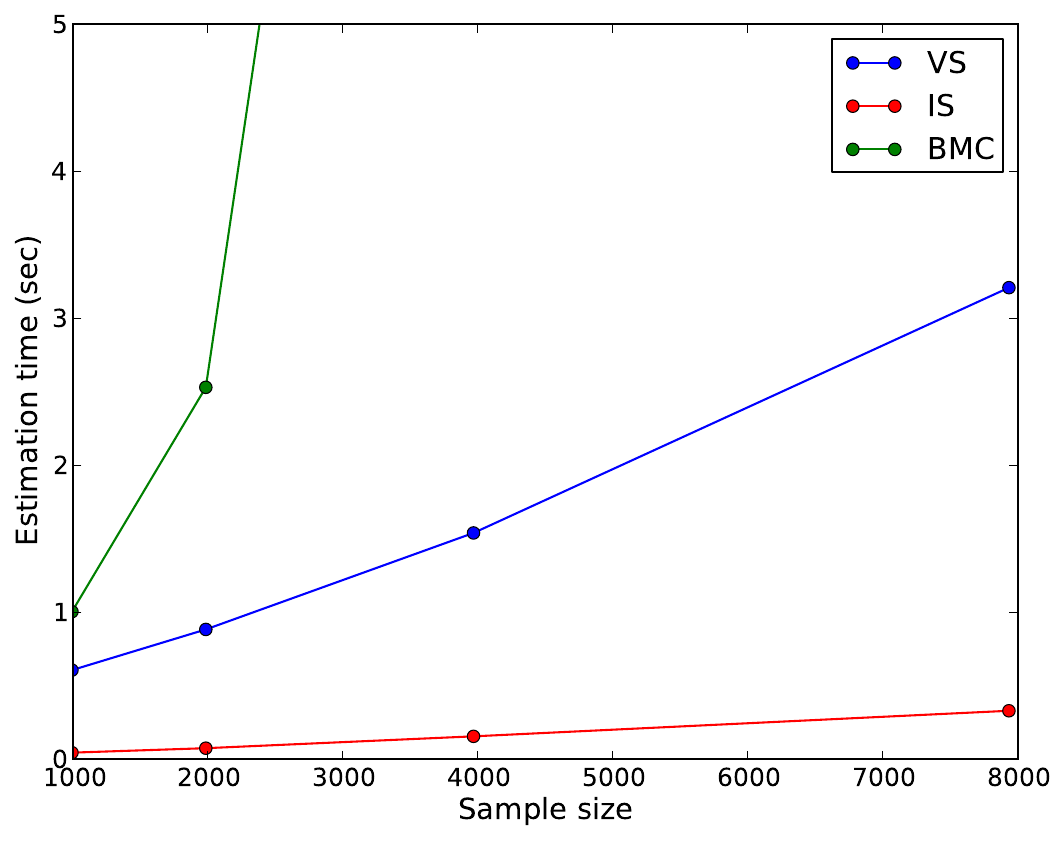}
  \end{center}
  \caption{Estimation time (excluding sampling time) for VS, IS and
    BMC in dimension 30.}
  \label{fig:timing}
\end{figure}

These findings must be tempered by the fact that VS requires more
computation time than IS {\em for a given sample size}, as shown in
Figure~\ref{fig:timing}. The good news is that estimation time scales
linearly with sample size similarly to IS, but unlike BMC where it
scales quadratically. As we will demonstrate now, one may actually
save considerable computation time by running VS on a moderate size
sample rather than running IS on a larger sample for equivalent
accuracy.

\subsection{Logistic regression}
\label{sec:logistic}

We here investigate VS and other methods to make Bayesian inference in
logistic regression \cite{Rasmussen-06}. In this case, the
unnormalized target distribution reads $ p(\x) = p_0(\x) \prod_{i=1}^M
\sigma(y_i \mathbf{a}_i^\top \x) $, where $y_i\in \{-1,1\}$ are binary
measurements in number $M$, $\sigma(x) = 1/(1 + e^{-x})$ is the
logistic function, $\mathbf{a}_i$ are known attribute vectors, $p_0$
is a prior, and $\x$ is an unknown vector of coefficients.

Quantities of special interest are the posterior mean and possibly the
posterior variance of $\x$ to make MAP or averaged predictions
\cite{Rasmussen-06}. It is customary to approximate these using the
Laplace method, which is fast but usually inaccurate. In the case of
probit regression, where $\sigma$ is replaced with the standard normal
cumulative distribution, there exists an instance of the EP~algorithm
with fully explicit updating rules (hereafter referred to as
EP-probit) that generally yields better results
\cite{Rasmussen-06}. Whilst explicit EP does not exist for logistic
regression, one may resort to the variational bound Gaussian
approximation proposed by Jaakkola \& Jordan \cite{Jaakkola-97},
denoted VB-logistic in the following, which minimizes an upper bound
of the exclusive KL divergence. A related variational bound method
using a mean-field approximation is discussed
in \cite{Knowles-11}. Other structured variational methods such as
Power~EP \cite{Minka-05} have the potential to work for logistic
regression but have not been reported so far. In \cite{Girolami-11},
several MCMC methods are compared against benchmark logistic
regression problems, among which the best performers tend to have high
computational cost.

The strategy we propose here is to use VS with sampling kernel $\pi$
chosen as the Laplace approximation to $p$.  We report below results
on several datasets from the UCI Machine Learning Repository
\cite{Bache-13} with attribute dimension ranging from $d=4$ to $34$,
where VS is compared to IS and BMC in the same sampling scenario, as
well as to EP-probit and VB-logistic.

In each experiment, a constant attribute was added to account for
uneven odds. All attributes were conventionally normalized to unit
Euclidean norm and a zero-mean Gaussian prior with large isotropic
variance $10^5$ was used. We evaluated VS, IS and BMC on random
samples of increasing sizes $N$ corresponding to multiples
$2,4,8,16,\ldots $ of the number of Gaussian parameters,
$n=(d+2)(d+1)/2$, provided that the overall computation time,
including Laplace approximation (which took only 10--50 milliseconds),
sampling and moment estimation, was below $10$~seconds. All
simulations satisfying this criterion were repeated $250$ times. BMC
was implemented using an isotropic Gaussian correlation function with
scale parameter tuned empirically to yield roughly optimal performance
rather than optimized by marginal likelihood maximization
\cite{Rasmussen-03}. This was done to keep computation time comparable
with both IS and VS.

Ground truth parameters were estimated using IS with samples of size
$10^7$. Excess KL divergences as defined in
Section~\ref{sec:variational_sampling} are reported in
Table~\ref{tab:logistic_errors} for each method. For VS, IS and BMC,
the reported value corresponds to the largest sample size that enabled
less than 10~seconds computation.  While both deterministic
variational methods EP-probit and VB-logistic ran fast (converging in
1--2 seconds), they both performed poorly in logistic regression.
This could be expected for EP-probit as it relies on a probit model
and indeed turned out to be the best performer for probit
regression. VB-logistic, however, was designed for logistic regression
but only works well in dimension $d=1$ in our experience.

\begin{figure}[!ht]
  \begin{center}
    \includegraphics[width=.5\textwidth]{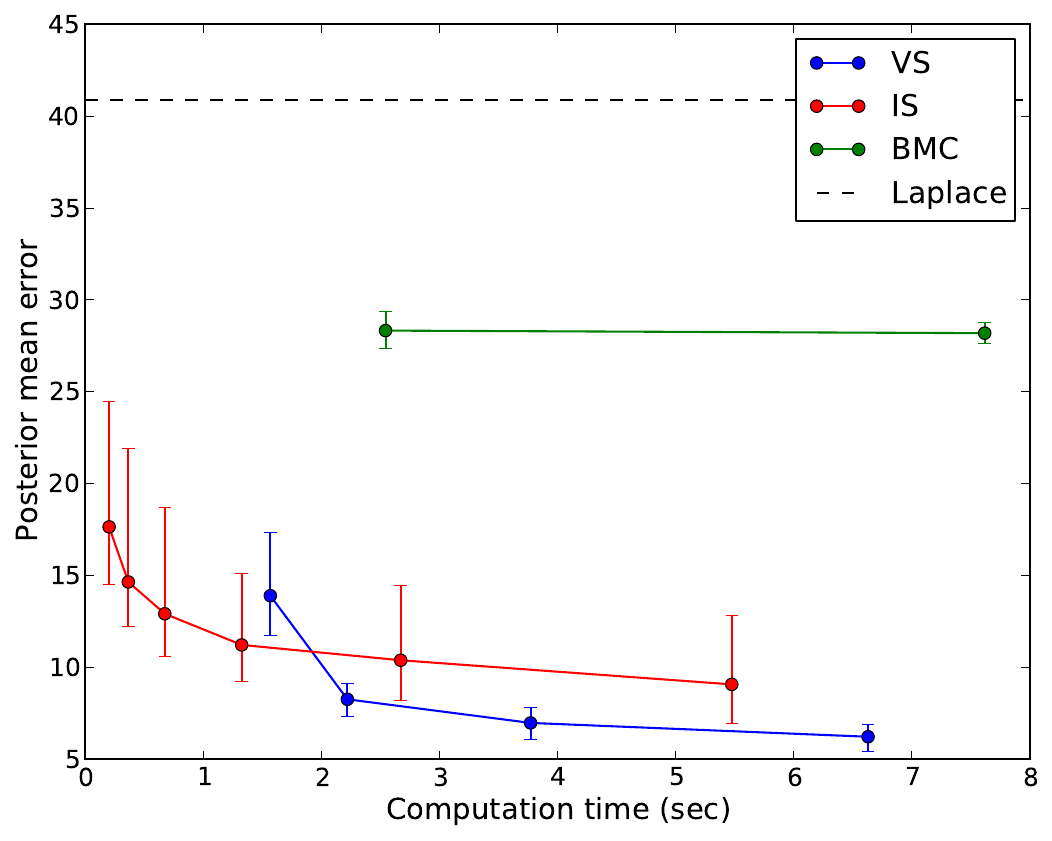}
    \includegraphics[width=.49\textwidth]{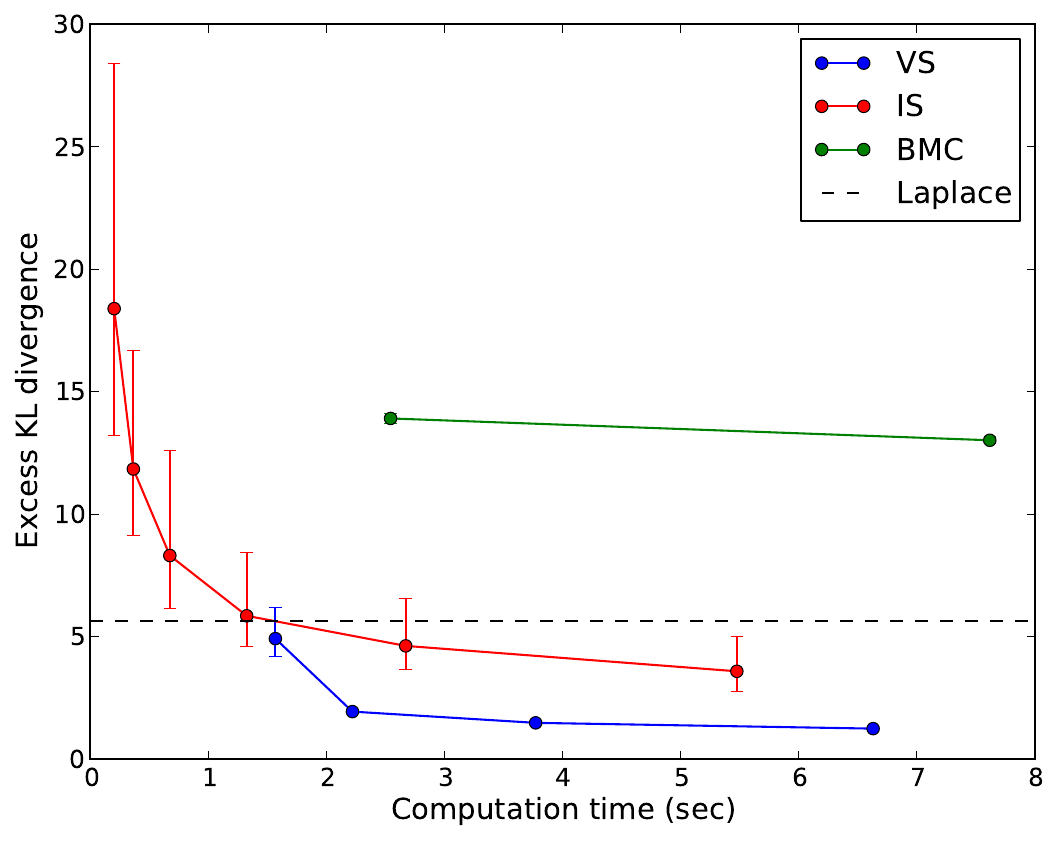}
\end{center}
\caption{Estimation errors in logistic regression on the
  ``Ionosphere'' UCI dataset. Left, errors on the posterior mean
  (Euclidean distances). Right, excess KL divergences. Dots and
  vertical segments respectively represent median errors and
  inter-quartile ranges across 250 sampling trials. Errors for
  EP-probit and VB-logistic are out of the represented range.}
  \label{fig:logistic_errors}
\end{figure}

Overall, logistic regression results show that VS was the only method
to clearly improve over the Laplace method in all datasets. VS proved
both more accurate and more precise than IS and BMC, while IS was more
accurate than BMC but generally less
precise. Figure~\ref{fig:logistic_errors} shows VS, IS and BMC errors
as decreasing functions of computation time, and confirms that VS has
the potential to massively overcome its lower computational efficiency
at given sample size compared to IS. Note that, since sampling time is
proportional to the number of measurements $M$, the computational
overhead of VS relative to IS is a decreasing function of $M$. An
additional observation is that IS and BMC tended to provide more
accurate posterior mean estimates than the Laplace method, but worse
variance estimates except for the ``Haberman'' low-dimensional
dataset.

These results suggest VS as a fast and accurate method for Bayesian
logistic regression. It is perhaps less useful in this implementation
for probit regression where, as shown by
Table~\ref{tab:logistic_errors}, VS still performed well and compared
favorably with IS and BMC but was less accurate than EP-probit within
10~seconds computation time. Since EP does not exactly minimize the
KL~divergence \cite{Minka-01,Minka-05}, VS could beat the EP~solution
using larger samples, as a consequence of
Proposition~\ref{prop:central_limit}, however such refinement might be
considered too costly in practice.

\begin{table}[!h]
  \begin{center}
    {\footnotesize
    \begin{tabular}{|c|c|c|c|c|c|c|c|c|}
      \hline
      Dataset & $d$ & $M$ & Model & EP-probit & VB-logistic & VS & IS & BMC \\  
      \hline\hline
      \multirow{2}{*}{Haberman} & 
      \multirow{2}{*}{4} & \multirow{2}{*}{306} & 
      Logistic & 
      $1941.9$ & $88.91$ & $0.001 \pm 0.000$ & $0.007 \pm 0.004$ & $0.015 \pm 0.007$\\
      & & & Probit & 
      $0.0005$ & $4811.11$ & $0.001 \pm 0.001$ & $0.157 \pm 0.074$ & $0.025 \pm 0.021$\\
      \hline
      \multirow{2}{*}{Parkinsons \cite{Little-07}} & 
      \multirow{2}{*}{23} & \multirow{2}{*}{195} &
      Logistic & 
      $13.22$ & $295.21$ & $0.12 \pm 0.01$ & $0.24 \pm 0.09$ & $1.21 \pm 0.05$\\
      & & & Probit & 
      $0.05$ & $129.90$ & $0.10 \pm 0.01$ & $0.32 \pm 0.13$ & $1.04 \pm 0.04$\\
      \hline
      \multirow{2}{*}{Ionosphere} & 
      \multirow{2}{*}{33} & \multirow{2}{*}{351} & 
      Logistic & 
      $17.07$ & $139.82$ & $0.22 \pm 0.02$ & $0.69 \pm 0.40$ & $2.30 \pm 0.05$\\
      & & & Probit & 
      $0.06$ & $58.60$ & $0.18 \pm 0.02$ & $0.56 \pm 0.27$ & $3.92 \pm 0.05$\\
      \hline
      Prognostic & 
      \multirow{2}{*}{34} & \multirow{2}{*}{194} & 
      Logistic & 
      $6.26$ & $24.49$ & $0.61 \pm 0.04$ & $0.83 \pm 0.22$ & $1.29 \pm 0.04$\\
      Breast Cancer & & & Probit & 
      $0.01$ & $26.10$ & $0.12 \pm 0.01$ & $0.48 \pm 0.20$ & $1.62 \pm 0.06$\\
      \hline
    \end{tabular}
    }
  \end{center}
  \caption{Excess KL divergences (divided by corresponding excess KL
    divergences of the Laplace method) in logistic and probit
    regression on UCI datasets: mid-hinges and inter-quartile ranges
    for several inference methods restricted to 10~seconds
    computation. Values below~1 indicate higher accuracy than the
    Laplace method.}
  \label{tab:logistic_errors}
\end{table}

\section{Conclusion}
\label{sec:conclusion}

In summary, VS is an asymptotically exact moment matching technique
that relies on few assumptions regarding the target distribution and
can provide an efficient alternative to the Laplace method as well as
conventional sampling-based and variational inference methods. VS may
be viewed as a {\em nonlinear} integration rule that is exact for
Gaussian functions (or, more generally, arbitrary exponential
families). Although the VS implementation tested in this paper uses
random sampling, we stress that VS is not intrinsically a Monte Carlo
method as it can be constructed from any integration rule.  Monte
Carlo VS was shown to work well in problems of moderate dimension (up
to~34 in the presented logistic regression examples) for which a
reasonable initial approximation to the target distribution was
available and was used as a sampling distribution. There are many
Bayesian inference problems where such an initial approximation can be
obtained efficiently using the Laplace method or other analytical
methods.

Extension to high dimension is challenging if only because the number
of free parameters in a full Gaussian model scales quadratically with
dimension. We see different potential ways to overcome this
difficulty. One is to use a sparse Gaussian approximation, such as one
with diagonal covariance model for which the number of parameters
$n=2d+1$ reduces to a linear function of dimension. This can help if
one is interested in marginal moments only, but can also slow down
convergence if strong component-wise correlations exist. Another
possible workaround is to try and factorize the target distribution
into a product of elementary distributions that each involve few
variables, and use VS as a building block within an EP-like algorithm
\cite{Minka-01,Minka-05} to repeatedly perform factor-wise
approximations if those cannot be computed analytically. Finally, a
more general approach is to use stochastic gradient descent methods
\cite{Bottou-08} to substitute quasi-Newton minimization in large
dimensional problems in order to cut down computational complexity and
memory load.


\end{document}